\documentclass[pra,showpacs,twocolumn]{revtex4-1}
\usepackage{amsfonts}
\usepackage{amsmath}
\usepackage{amssymb,amscd}
\usepackage{psfrag}
\usepackage{graphicx}
\usepackage{braket}
\usepackage[dvips]{epsfig}
\usepackage{epsfig}
\usepackage{subfigure}
\usepackage{color}
\usepackage{amssymb}

\begin{document}

\title{Ultimate precision limit of SU(2) and SU(1,1) interferometers in
noisy metrology}
\author{Jie Zeng$^{1}$}
\author{Dong Li$^{2,3}$}
\author{L. Q. Chen$^{1,4}$}
\email{lqchen@phy.ecnu.edu.cn}
\author{Weiping Zhang$^{4,5,6,7}$}
\author{Chun-Hua Yuan$^{1,4}$}
\email{chyuan@phy.ecnu.edu.cn}

\address{$^1$State Key Laboratory of Precision Spectroscopy, Quantum Institute for Light and Atoms,
	Department of Physics, East China Normal University, Shanghai 200062, China}
\address{$^2$Microsystems and Terahertz Research Center, China Academy of
Engineering Physics, Chengdu, Sichuan 610200, China}
\address{$^3$Institute
of Electronic Engineering, China Academy of Engineering Physics, Mianyang,
Sichuan 621999, China}
\address{$^4$Shanghai Branch, Hefei National Laboratory, Shanghai 201315,
China}
\address{$^5$Department of Physics, Shanghai Jiao Tong University, and
Tsung-Dao Lee Institute, Shanghai 200240, China}
\address{$^6$Shanghai
Research Center for Quantum Sciences, Shanghai 201315, China}
\address{$^7$Collaborative Innovation Center of Extreme Optics, Shanxi
University, Taiyuan, Shanxi 030006, China}

\begin{abstract}
The quantum Fisher information (QFI) in SU(2) and SU(1,1) interferometers
was considered, and the QFI-only calculation was overestimated. In general,
the phase estimation as a two-parameter-estimation problem, and the quantum
Fisher information matrix (QFIM) is necessary. In this paper, we
theoretically generalize the model developed by Escher \textit{et al}
[Nature Physics 7, 406 (2011)] to the QFIM case with noise and study the ultimate
precision limits of SU(2) and SU(1,1) interferometers with photon losses because photon losses as a very usual noise may happen to the phase measurement process. Using coherent state $\otimes $ squeezed vacuum state as a specific example, we numerically analyze the variation of the overestimated QFI with the loss
coefficient or splitter ratio, and find its disappearance and recovery phenomenon. By adjusting the splitter ratio and loss coefficient the optimal sensitivity is obtain, which is beneficial to quantum precision measurement in a lossy environment.
\end{abstract}

\date{\today }
\maketitle

\section{Introduction}

In quantum metrology, the Mach--Zehnder interferometer (MZI) has been
exploited as a generic tool to realize precise measurement of phase. The
sensitivity of interferometer measurements is restricted by the shot-noise
limit, or the standard quantum limit (SQL) with respect to classic
resources. Therefore, how to improve the sensitivity of interferometers has
been received a lot of attention in recent years \cite%
{Helstrom67,Holevo82,Caves81,Braunstein94,Braunstein96,Lee02,Giovannetti06,Zwierz10,Giovannetti04,Giovannetti11}%
, in which many quantum techniques have been utilized to improve measurement
precision than purely classical approaches.

Yurke et al. \cite{Yurke86} theoretically introduced the SU(1,1)
interferometer using an active nonlinear beam splitter (NBS) in place of a
passive linear beam splitter (LBS) for wave splitting and recombination. It
is called the SU(1,1) interferometer because it is described by the SU(1,1)
group, as opposed to SU(2) group for BSs. The NBS can be provided by the
optical parameter amplifier process or the four-wave mixing process. Due to
the quantum destructive interference in the SU(1,1) interferometer, the
noise accompanied by the amplification of the signal can revert to the level
of input. Benefiting from that, the signal-to-noise ratio improves. Because
it can be used to improve measurement sensitivity, this type of
interferometers has received extensive attention both experimentally \cite%
{Jing11,Hudelist14,Chen15,Qiu16,Linnemann16,Lemieux16,Manceau17,Anderson17,Gupta17,Du18,Frascella19,Prajapati19,Du22}
and theoretically \cite%
{Plick10,Ou12,Marino12,Li14,Gabbrielli15,Ma15,Chen16,Sparaciari16,Li16,Hu16,Gong17,Giese17,Li18,Guo18,Hu18,Ma18,Michael19,Gao20,Ou20,Gao22, Liang22}%
.

In the presence of environment noise, the measurement precision of
interferometers will be reduced because there are inevitable interactions
with the surrounding environment, which have been studied by many
researchers \cite{Dem09,Escher12,Dem12,Yue14,Berry13, Chaves, Dur,
Kessler,Brivio,Alipour,Escher11,Genoni11,Genoni12,Feng14}. For
interferometers, the photon losses is a typical decoherence process which
should be taken into account. The general frame for estimating the ultimate
precision limite in the presence of photo loss has been analyzed \cite%
{Dem09,Escher11,Dem12,Yue14}, where this decoherence process can be
described by a set of Kraus operators, and the corresponding lower bounds in
quantum metrology is given by the quantum Cra\'{m}er-Rao bound (QCRB) usage
of quantum Fisher information (QFI) \cite{Braunstein94,Braunstein96}. It
establishes the best precision that can be attained with a given quantum
probe \cite{Toth, PezzBook, Demkowicz,Wang,Monras, Pinel, Liu,
Gao,Jiang,Safranek,Sparaciari15,Gao21,Chang22}.

For the SU(2) and SU(1,1) interferometers without losses, the QFIs of
unknown phase shifts only in the single arm case or in the two-arm case have
been investigated, where the results of QFI-only calculations will be
overestimated because many measurement processes do not include phase
reference (or phase-averaging) \cite{Jarzyna,Gong17CH,You19,Takeoka}. For
the unknown phase shifts in the two-arm case, the phase estimation is indeed
a two-parameter estimation problem. Although, the results of phase-averaging
method and multiparameter estimation are the same for some particular inputs
\cite{Takeoka,You19}. In general, the calculation of the quantum Fisher
information matrix (QFIM) is necessary when the unknown phase shifts are
applied in both arms \cite{Takeoka,You19,Lang1, Lang2, Ataman, Zhong2}. In
the presence of losses, the QFIM of SU(2) and SU(1,1) interferometers and
corresponding QCRBs are worth studying because there are inevitable
interactions with the surrounding environment in the measurement process.

In this work, we theoretically study the ultimate precision limit of SU(2)
and SU(1,1) interferometers in the presence of losses. In order to avoid
overestimation of QFI-only calculation, we give the phase sensitivity of
SU(2) and SU(1,1) interferometers with the QFIM approach for the unknown
phase shifts in the two-arm case. When inputs are arbitrary pure states, the
ultimate precision limit for the SU(2) and SU(1,1) interferometers with
optical losses in single arm and two arms are given. Taking the coherent
states $\otimes $ squeezed vacuum input as a specific example, we
numerically calculate and compare the sensitivities between the
single-parameter estimation and two-parameter estimation.

The paper is organized as follows. In Sec.~II, we firstly review the
QFIM of the SU(2) and SU(1,1) interferometers in the lossless case, and
derive the QFIM of them with noise with the method proposed by Escher et al.
\cite{Escher11}. In Sec. III and IV, we respectively investigate the QFIMs
in the case of optical losses in single arm and two arms. In Sec. V, we
numerically calculate and compare the QCRBs between the single-parameter
estimation and two-parameter estimation. Finally, we conclude with a summary
of our results.

\begin{figure}[t]
\centerline{\includegraphics[width=0.4\textwidth,angle=0]{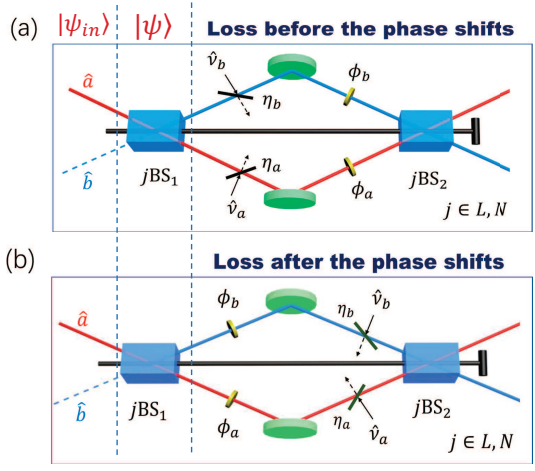}}
\caption{Lossy interferometer model. The losses in the interferometer are
modeled by adding fictitious beam splitters (a) before and (b) after the
phase shifts $\protect\phi _{a}$ and $\protect\phi _{b}$. The beam splitting
is the passive linear beam splitters (LBSs) or the active nonlinear beam
splitters (NBSs), corresponding the SU(2) interferometers or SU(1,1)
interferometers, respectively. $\hat{v}_{a}$ ($\hat{v}_{b}$) is the vacuum,
and $\protect\eta_{a}$ ($\protect\eta_{b}$) is transmission rate. In
realistic systems the photon losses are distributed throughout the arms of
the interferometer, not an event concentrated either before or after the
phase, which is handled with a distribution parameter.}
\label{fig1}
\end{figure}

\section{ QFIM of interferometers}

\label{sec:QOI}

\subsection{Ideal QFIM}

\label{sec:IQ} We consider a general interferometer, as shown in Fig.~\ref%
{fig1}, where the beam splitter can be either linear or nonlinear,
corresponding SU(2) and SU(1,1) interferometers, respectively. Consider the
initial input state $|\psi _{in}\rangle $. After the first LBS or NBS, the
state is labeled by $|\psi \rangle $. The two beams sustain phase shifts,
i.e., mode $a$ undergoes a phase shift of $\phi _{a}$\ and mode $b$
undergoes a phase shift of $\phi _{b}$, and the state is $\left\vert \psi
_{\phi }\right\rangle $. Then the phase transformation is written as
\begin{equation}
\hat{U}_{\phi }^{ab}=\exp [i\phi _{a}\hat{n}_{a}+i\phi _{b}\hat{n}_{b}]=\exp
(i\phi _{+}\hat{g}_{+})\exp (i\phi _{-}\hat{g}_{-}),
\end{equation}%
where $\phi _{\pm }=\phi _{a}\pm \phi _{b}$ and $g_{\pm }=(\hat{n}_{a}\pm
\hat{n}_{b})/2$ with $\hat{n}_{a}=\hat{a}^{\dag }\hat{a}$ and $\hat{n}_{b}=%
\hat{b}^{\dag }\hat{b}$.

The QFI is the intrinsic information in the quantum state and is not related
to the actual measurement procedure, which gives an upper limit to the
precision of quantum parameter estimation. However, the QFI-only calculation
was overestimated. In general, the phase estimation as a two-parameter
estimation problem such as $\phi _{a}$ and $\phi _{b}$ (or $\phi _{+}$ and $%
\phi _{-}$), and the quantum Fisher information matrix (QFIM) is necessary.

In the SU(2) and SU(1,1) interferometers, the estimated phases are $\phi
_{-} $ and $\phi _{+}$, respectively \cite{Kok2010}. For the estimation of $%
\phi _{+}$ and $\phi _{-}$ we can use the method of QFIM, which is given by
a two-by-two matrix
\begin{equation}
\mathcal{F}_{QFIM}=\left[
\begin{array}{cc}
F_{++} & F_{+-} \\
F_{-+} & F_{--}%
\end{array}%
\right] ,
\end{equation}%
where $F_{ij}=4(\langle \hat{g}_{i}\hat{g}_{j}\rangle -\langle \hat{g}%
_{i}\rangle \langle \hat{g}_{j}\rangle )$ ($i$, $j=+,-$), $\langle \cdot
\rangle $ denotes the average value $\left\langle \psi _{\phi }\right\vert
\cdot |\psi _{\phi }\rangle $ as shown in Fig.~\ref{fig1}. The estimation of
$\phi _{-}$ and $\phi _{+}$ are respectively given by%
\begin{equation}
\Delta ^{2}\phi _{-}\geq \frac{1}{\mathcal{F}_{QFIM_{-}}}\text{, \ }\Delta
^{2}\phi _{+}\geq \frac{1}{\mathcal{F}_{QFIM_{+}}},
\end{equation}%
where%
\begin{eqnarray}
\mathcal{F}_{QFIM_{-}} &=&F_{--}-\Delta F_{-}\text{, }\mathcal{F}%
_{QFIM_{+}}=F_{++}-\Delta F_{+},  \notag \\
\Delta \mathcal{F}_{-} &=&\frac{F_{+-}F_{-+}}{F_{++}}\text{, }\Delta
\mathcal{F}_{+}=\frac{F_{+-}F_{-+}}{F_{--}}.
\end{eqnarray}%
When $\Delta \mathcal{F}_{\pm }\neq 0$, $\Delta \mathcal{F}_{\pm }$\ are the
overestimated Fisher information from QFI-only calculation. If one ignores
the nondiagonal terms, i.e., implicity assuming that $\phi _{+}$ ($\phi _{-}
$)\ is known a \textit{priori}, one could use the diagonal terms $F_{--} $ ($%
F_{++})$ to indicate\ the QFI of the single-parameter $\phi _{-}$ ($\phi
_{+} $) estimation, where the misleading overestimates the precision limit
\cite{Gong17,Takeoka,You19}.

For a two-input-port interferometer, the matrix elements $F_{ij}$ are given
by
\begin{eqnarray}
F_{++} &=&\langle \Delta ^{2}\hat{n}_{a}\rangle +\langle \Delta ^{2}\hat{n}%
_{b}\rangle +2Cov[n_{a},n_{b}],  \notag \\
F_{--} &=&\langle \Delta ^{2}\hat{n}_{a}\rangle +\langle \Delta ^{2}\hat{n}%
_{b}\rangle -2Cov[n_{a},n_{b}],  \notag \\
F_{+-} &=&F_{-+}=\langle \Delta ^{2}\hat{n}_{a}\rangle -\left\langle \Delta
^{2}\hat{n}_{b}\right\rangle ,
\end{eqnarray}%
where $\langle \Delta ^{2}\hat{n}_{i}\rangle =\langle \psi |\hat{n}%
_{i}^{2}|\psi \rangle -\langle \psi |\hat{n}_{i}|\psi \rangle ^{2}$ $(i=a$, $%
b)$, and $Cov[\hat{n}_{a},\hat{n}_{b}]=\langle \psi |\hat{n}_{a}\hat{n}%
_{b}|\psi \rangle -\langle \psi |\hat{n}_{a}|\psi \rangle \langle \psi |\hat{%
n}_{b}|\psi \rangle $. $Cov[\hat{n}_{a},\hat{n}_{b}]$ is the covariance of
two-mode field to describe the intermode correlation. After calculation, the
ultimate $\mathcal{F}_{QFIM_{-}}$\ and $\mathcal{F}_{QFIM_{+}}$ are\
rewritten as
\begin{eqnarray}
\mathcal{F}_{QFIM_{-}} &=&4\frac{\langle \Delta ^{2}\hat{n}_{a}\rangle
\langle \Delta ^{2}\hat{n}_{b}\rangle -Cov[n_{a},n_{b}]^{2}}{\langle \Delta
^{2}\hat{n}_{a}\rangle +\langle \Delta ^{2}\hat{n}_{b}\rangle
+2Cov[n_{a},n_{b}]}, \\
\mathcal{F}_{QFIM_{+}} &=&4\frac{\langle \Delta ^{2}\hat{n}_{a}\rangle
\langle \Delta ^{2}\hat{n}_{b}\rangle -Cov[n_{a},n_{b}]^{2}}{\langle \Delta
^{2}\hat{n}_{a}\rangle +\langle \Delta ^{2}\hat{n}_{b}\rangle
-2Cov[n_{a},n_{b}]}.
\end{eqnarray}%
and the overestimated Fisher information are cast into
\begin{eqnarray}
\Delta \mathcal{F}_{-} &=&\frac{\left( \langle \Delta ^{2}\hat{n}_{a}\rangle
-\langle \Delta ^{2}\hat{n}_{b}\rangle \right) ^{2}}{\langle \Delta ^{2}\hat{%
n}_{a}\rangle +\langle \Delta ^{2}\hat{n}_{b}\rangle +2Cov[n_{a},n_{b}]},
\label{DFM} \\
\Delta \mathcal{F}_{+} &=&\frac{\left( \langle \Delta ^{2}\hat{n}_{a}\rangle
-\langle \Delta ^{2}\hat{n}_{b}\rangle \right) ^{2}}{\langle \Delta ^{2}\hat{%
n}_{a}\rangle +\langle \Delta ^{2}\hat{n}_{b}\rangle -2Cov[n_{a},n_{b}]}.
\label{DFP}
\end{eqnarray}%
This is a general expression, which is dependent on the form of the input
state, and beam splitting transformation such as BS or NBS.

\begin{figure}[t]
\centerline{\includegraphics[width=0.5\textwidth,angle=0]{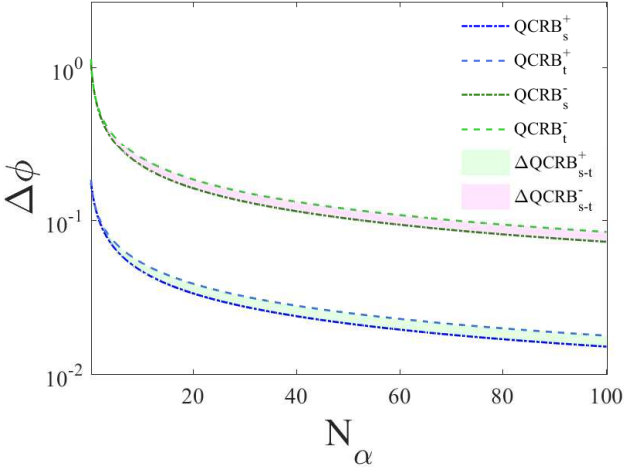}}
\caption{Phase sensitivity $\Delta\protect\phi$ of SU(2) interferometers
(top part) and SU(1,1) interferometers (bottom part) as a function of $N_{%
\protect\alpha}$ with $r=1.5$, $T=0.7$, and $G_{1}=1.2$. QCRB$^{k}_{p}$: $%
k\in$[+, SU(1,1) case; -, SU(2) case], $p\in$[s, QFI case; t, QFIM case].
The shaded parts represent the difference in phase sensitivities due to
overestimated QFI.}
\label{fig2}
\end{figure}

Using the linear correlation coefficient $J$ to describe the intermode
correlations \cite{Gerry05}, and the Mandel $Q$ parameter to describe the
intramode \cite{Mandel95}, which are given by
\begin{equation}
J=\frac{Cov[\hat{n}_{a},\hat{n}_{b}]}{\langle \Delta \hat{n}_{a}\rangle
\langle \Delta \hat{n}_{b}\rangle },\text{ }Q_{i}=\frac{\left\langle \Delta
^{2}\hat{n}_{i}\right\rangle -\langle \hat{n}_{i}\rangle }{\langle \hat{n}%
_{i}\rangle },
\end{equation}%
where $\langle \Delta \hat{n}_{i}\rangle =\sqrt{\langle \Delta ^{2}\hat{n}%
_{i}\rangle }$ $(i=a,b)$. Then, the ultimate $F_{QFIM_{-}}$\ and $%
F_{QFIM_{+}}$ are rewritten as
\begin{eqnarray}
\mathcal{F}_{QFIM_{-}} &=&\frac{4\sqrt{\left\langle \hat{n}_{a}\right\rangle
\left\langle \hat{n}_{b}\right\rangle (Q_{a}+1)(Q_{b}+1)}(1-J^{2})}{\sqrt{%
\frac{\left\langle \hat{n}_{b}\right\rangle (Q_{b}+1)}{\left\langle \hat{n}%
_{a}\right\rangle (Q_{a}+1)}}+\sqrt{\frac{\left\langle \hat{n}%
_{a}\right\rangle (Q_{a}+1)}{\left\langle \hat{n}_{b}\right\rangle (Q_{b}+1)}%
}+2J},\text{ } \text{ } \\
\mathcal{F}_{QFIM_{+}} &=&\frac{4\sqrt{\left\langle \hat{n}_{a}\right\rangle
\left\langle \hat{n}_{b}\right\rangle (Q_{a}+1)(Q_{b}+1)}(1-J^{2})}{\sqrt{%
\frac{\left\langle \hat{n}_{b}\right\rangle (Q_{b}+1)}{\left\langle \hat{n}%
_{a}\right\rangle (Q_{a}+1)}}+\sqrt{\frac{\left\langle \hat{n}%
_{a}\right\rangle (Q_{a}+1)}{\left\langle \hat{n}_{b}\right\rangle (Q_{b}+1)}%
}-2J}.\text{ } \text{ }
\end{eqnarray}

If $\langle \Delta ^{2}\hat{n}_{a}\rangle =\langle \Delta ^{2}\hat{n}%
_{b}\rangle $, the overestimated Fisher information $\Delta \mathcal{F}_{\pm
}$ will be $0$, and the results of phase estimation $\phi _{\pm }$ are from
the single-parameter QFI. Furthermore, considering $\left\langle \hat{n}%
_{a}\right\rangle =\left\langle \hat{n}_{b}\right\rangle =\bar{n}/2$, $\bar{n%
}$ is the average number of photons in the probe state, the results $%
\mathcal{F}_{QFIM_{\pm }}$ are reduced to
\begin{eqnarray}
\mathcal{F}_{QFIM_{-}} &=&\mathcal{F}_{QFI_{-}}=\bar{n}(Q+1)(1-J), \\
\mathcal{F}_{QFIM_{+}} &=&\mathcal{F}_{QFI_{+}}=\bar{n}(Q+1)(1+J).
\end{eqnarray}%
They are the same as that of the SU(2) \cite{Sahota15} and SU(1,1) \cite%
{Gong17}, respectively. The intermode correlation $J$ ranges between $-1$
and $1$, which contribute at most a factor of $2$ improvement in the QFI.
The photon statistics within each arm of the interferometer are
super-Poissonian and sub-Poissonian for $Q>0$ and $-1<Q<0$, respectively.
Larger quantum enhancement originates from the photon variances within each
arm of the interferometers \cite{Sahota15,Gong17}.

In the case of $\langle \Delta ^{2}\hat{n}_{a}\rangle \neq \langle \Delta
^{2}\hat{n}_{b}\rangle $, for the SU(2) interferometers it exists when the
beam splitting ratio of the first LBS is not 50:50 without losses \cite%
{Jarzyna}. However, in the presence of losses the optimal splitting ratio of
LBS for the SU(2) interferometers is no longer 50:50, leading to $\langle
\Delta ^{2}\hat{n}_{a}\rangle \neq \langle \Delta ^{2}\hat{n}_{b}\rangle $
\cite{Huang22,Cooper11}. For the SU(1, 1) interferometers, $\langle \Delta
^{2}\hat{n}_{a}\rangle \neq \langle \Delta ^{2}\hat{n}_{b}\rangle $ exists
as long as that the input states of the two inputs are not the same \cite%
{Yu22}. Therefore, it is very necessary to estimate the phase sensitivity
with noise by using the method of QFIM.

\subsection{ Lossy QFIM}

In this section, the general model, developed by Escher \textit{et al}. \cite%
{Escher11} for open quantum system metrology, is extended to the QFIM case.

We begin with a system $S$ and consider it along with the environment $E$.
In general, the enlarged state of system and environment ($S+E$) evolves as%
\begin{equation}
|\Psi \rangle _{SE}=\hat{U}_{SE}(x_{+},x_{-})|\psi \rangle _{S}|0\rangle
_{E}=\sum_{l}\hat{\Pi}_{l}(x_{+},x_{-})|\psi \rangle _{S}|l\rangle _{E},
\end{equation}%
where $\hat{U}_{SE}(x_{+},x_{-})$ is the corresponding unitary operator, and
$|\psi \rangle _{S}$ is the initial state of the probe. $|0\rangle _{E}$\ is
the initial state of the environment, and $|l\rangle _{E}$\ are orthogonal
states of the environment $E$. $\hat{\Pi}_{l}(x_{+},x_{-})$ are $x_{+}$, and
$x_{-}$-dependent Kraus operators, which act on the system $S$. The
evolution of an open system can be described as%
\begin{eqnarray}
\rho _{S} &=&Tr_{E}\left[ \hat{U}_{SE}\rho _{SE}(0)\hat{U}_{SE}^{\dag }%
\right]  \notag \\
&=&\sum_{l}\hat{\Pi}_{l}(x_{+},x_{-})|\psi \rangle _{SS}\langle \psi |\hat{%
\Pi}_{l}^{\dag }(x_{+},x_{-}),
\end{eqnarray}%
where $\rho _{SE}(0)$ is the state of the systems and environment. For the
enlarged system-environment state, the QFI $\mathcal{C}_{QFIM}$\ is given by
\begin{equation}
\mathcal{C}_{QFIM}=\left[
\begin{array}{cc}
C_{++} & C_{+-} \\
C_{-+} & C_{--}%
\end{array}%
\right] ,
\end{equation}%
where
\begin{eqnarray}
C_{ij} &=&4[\left\langle \psi \right\vert _{S}\left\langle l\right\vert
_{E}\sum_{l}\frac{d\hat{\Pi}_{l}^{\dag }(x_{+},x_{-})}{d\phi _{i}}\frac{d%
\hat{\Pi}_{l}(x_{+},x_{-})}{d\phi _{j}}\left\vert \psi \right\rangle
_{S}\left\vert l\right\rangle _{E}  \notag \\
&&-\left\langle \psi \right\vert _{S}\left\langle l\right\vert _{E}\sum_{l}i%
\frac{d\hat{\Pi}_{l}^{\dag }(x_{+},x_{-})}{d\phi _{i}}\hat{\Pi}%
_{l}(x_{+},x_{-})\left\vert \psi \right\rangle _{S}\left\vert l\right\rangle
_{E}  \notag \\
&&\times \left\langle \psi \right\vert _{S}\left\langle l\right\vert
_{E}\sum_{l}-i\hat{\Pi}_{l}^{\dag }(x_{+},x_{-})\frac{d\hat{\Pi}%
_{l}(x_{+},x_{-})}{d\phi _{j}}\left\vert \psi \right\rangle _{S}\left\vert
l\right\rangle _{E},  \notag \\
i,j &\in &(+,-).
\end{eqnarray}%
The matrix elements $C_{ij}$ above can be expressed as%
\begin{eqnarray}
C_{ij} &=&4[\left\langle \psi \right\vert _{S}\left\langle l\right\vert _{E}%
\hat{H}_{ij}(\phi _{+},\phi _{-})\left\vert \psi \right\rangle
_{S}\left\vert l\right\rangle _{E}  \notag \\
&&-\left\langle \psi \right\vert _{S}\left\langle l\right\vert _{E}\hat{h}%
_{i}(\phi _{+},\phi _{-})\left\vert \psi \right\rangle _{S}\left\vert
l\right\rangle _{E}  \notag \\
&&\times \left\langle \psi \right\vert _{S}\left\langle l\right\vert _{E}%
\hat{h}_{j}^{\prime }(\phi _{+},\phi _{-})\left\vert \psi \right\rangle
_{S}\left\vert l\right\rangle _{E}],  \label{CQ}
\end{eqnarray}%
where
\begin{eqnarray}
\hat{H}_{ij}(\phi _{+},\phi _{-}) &=&\sum_{l}\frac{d\hat{\Pi}_{l}^{\dag
}(\phi _{+},\phi _{-})}{d\phi _{i}}\frac{d\hat{\Pi}_{l}(\phi _{+},\phi _{-})%
}{d\phi _{j}},  \notag \\
\hat{h}_{i}(\phi _{+},\phi _{-}) &=&i\sum_{l}\frac{d\hat{\Pi}_{l}^{\dag
}(\phi _{+},\phi _{-})}{d\phi _{i}}\hat{\Pi}_{l}(\phi _{+},\phi _{-}),
\notag \\
\hat{h}_{j}^{\prime }(\phi _{+},\phi _{-}) &=&-i\sum_{l}\hat{\Pi}_{l}^{\dag
}(\phi _{+},\phi _{-})\frac{d\hat{\Pi}_{l}(\phi _{+},\phi _{-})}{d\phi _{j}}.
\end{eqnarray}

By using QFIM method, similar to lossless case, we obtain the estimation of
the phase difference $\phi _{-}$ and phase sum $\phi _{+}$ in the presence
of losses, which is bounded by the QFIM:
\begin{eqnarray}
\mathcal{C}_{QFIM_{-}} &=&C_{--}-\Delta C_{-},  \label{CQFIM} \\
\mathcal{C}_{QFIM_{+}} &=&C_{++}-\Delta C_{+},  \label{CQFIMP}
\end{eqnarray}%
where
\begin{equation}
\Delta C_{-}=\frac{C_{+-}C_{-+}}{C_{++}}\text{, }\Delta C_{+}=\frac{%
C_{+-}C_{-+}}{C_{--}}.
\end{equation}%
$\Delta C_{\pm }$\ are the overestimated Fisher information from QFI-only
calculation in the presence of losses. Because the additional freedom
supplied by the environment should increase the QFI. Therefore, $\mathcal{C}%
_{QFIM_{+}}$ ($\mathcal{C}_{QFIM_{-}}$) should be larger or equal to $%
\mathcal{F}_{QFIM_{+}}$\ ($\mathcal{F}_{QFIM_{-}}$) . The relation between $%
\mathcal{F}_{QFIM_{\pm }}$\ and $\mathcal{C}_{QFIM_{\pm }}$ is found to be%
\begin{eqnarray}
\mathcal{F}_{QFIM_{i}} &=&\underset{\{\hat{\Pi}_{l}(x_{+},x_{-})\}}{\min }%
\mathcal{C}_{QFIM_{i}}\left[ |\psi \rangle ,\hat{\Pi}_{l}(x_{+},x_{-})\right]
,  \notag \\
i &\in &(+,-).
\end{eqnarray}

Usually, losses in the interferometers can be modeled by adding the
fictitious beam splitters, and the lossy evolution of the field in two arms
is described by the Kraus operator $\hat{\Pi}_{l}(\phi _{+},\phi _{-})$ with
considering the phase shift. In realistic systems the photon losses are
distributed throughout the arms of the interferometer, which is described by
the parameter $\gamma $, instead of simply inserting the fictitious beam
splitters before or after the phase shift.

Therefore, in the presence of losses the matrix elements of $\mathcal{F}%
_{QFIM_{\pm }}$ are worked in three steps: (1) substituting the $\hat{\Pi}%
_{l}(\phi _{+},\phi _{-})$ into Eq.~(\ref{CQ}), we can obtain the $C_{ij}$,
which is a function of $\gamma $; (2) minimizing $\mathcal{C}_{QFIM_{\pm }}$
by the parameters $\gamma $, the optimal $\gamma _{opt}^{\pm }$ is obtained;
(3) substituting $\gamma _{opt}^{\pm }$ into $\mathcal{C}_{QFIM_{\pm }}$,
the minimum $\mathcal{C}_{QFIM_{\pm }}^{opt}$ is obtain, i.e., $\mathcal{F}%
_{QFIM_{\pm }}$ is achieved.

Next, we applied this model to analyze the phase estimation in the presence
of losses in the case of in one arm and in two arms.

\begin{figure}[t]
\centerline{\includegraphics[width=0.5\textwidth,angle=0]{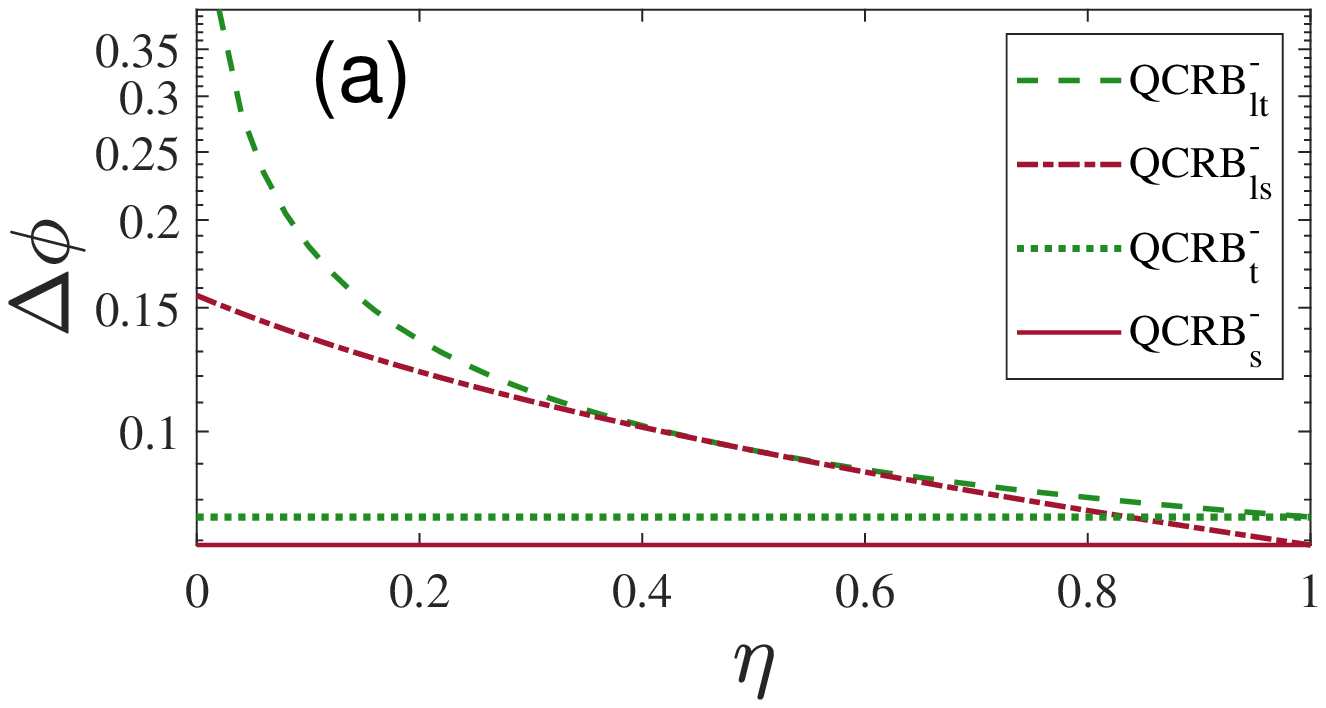}} %
\centerline{\includegraphics[width=0.5\textwidth,angle=0]{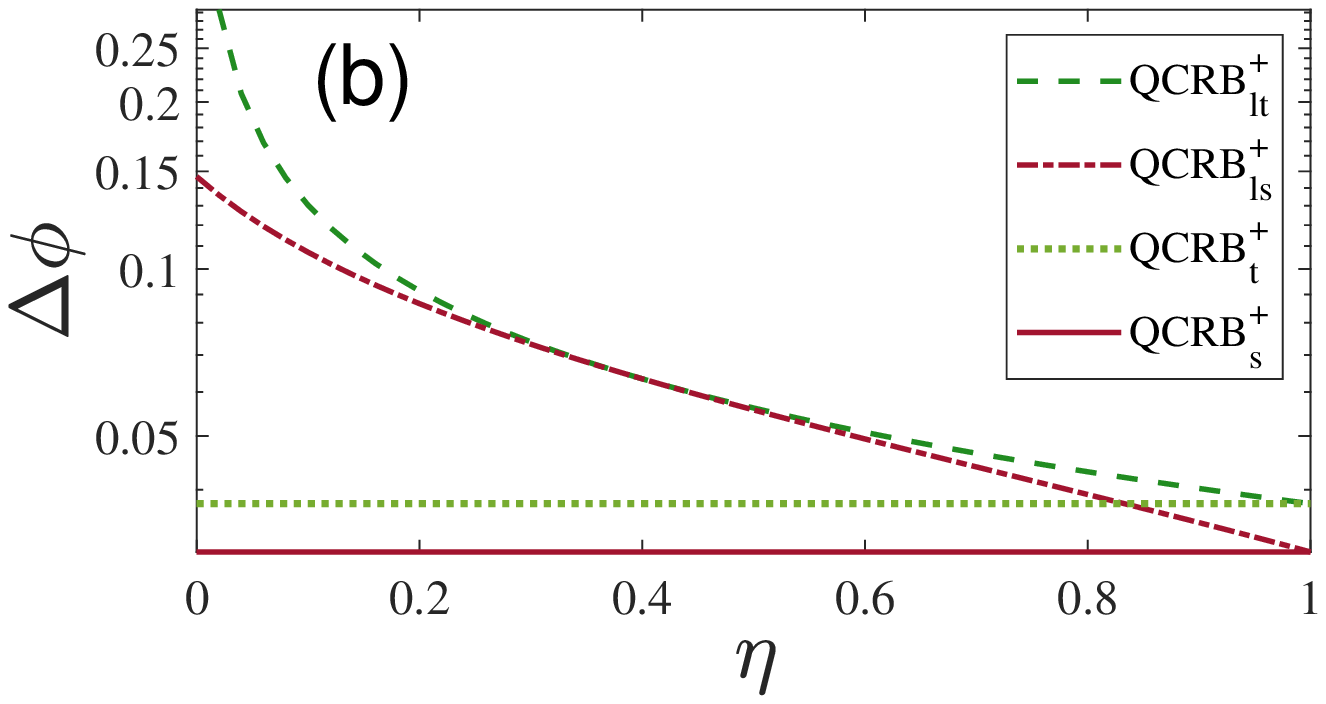}}
\caption{Phase sensitivities of (a) SU(2) interferometers and (b) SU(1,1)
interferometers versus photon loss coefficient $\protect\eta$ with
single-parameter estimation and two-parameter estimation with $\left |%
\protect\alpha \right | =10$, $r=0.5$ and $R/T=0.25$ in (a), $G_{1}=1.2$ in
(b). QCRB$^{k}_{p}$: $k\in$[+, SU(1,1) case; -, SU(2) case], $p\in$[s, QFI
case; ls, QFI case with loss; t, QFIM case; lt, QFIM case with loss].}
\label{fig3}
\end{figure}

\section{Losses in one arm}

Photon losses, a very usual noise, may happen at any stage of the phase process and is modeled
by the fictitious LBS introduced in the interferometer arms. Firstly, we consider the photon losses in just one of the two arms, for example arm $a$. A possible set of Kraus operators describing
the process without considering the phase shift is%
\begin{equation}
\hat{\pi}_{l_{a}}=\sqrt{\frac{(1-\eta _{a})^{l_{a}}}{l_{a}!}}\eta _{a}^{\hat{%
n}_{a}/2}\hat{a}^{l_{a}},
\end{equation}%
where $\eta _{a}$ quantifies the photon losses of arm $a$ ($\eta _{a}=1$,
lossless case; $\eta _{a}=0$, complete absorption).

When the photon losses before or after the phase shifts as shown in Fig. 1,
the Kraus operators $\hat{\pi}_{l_{a}}(\phi _{+},\phi _{-})$ including the
phase factor the general form ($\hbar =1$) is given by
\begin{eqnarray}
\hat{\Pi}_{la}(\varphi _{+},\varphi _{-}) &=&\sqrt{\frac{(1-\eta
_{a})^{l_{a}}}{l_{a}!}}e^{i\phi _{-}\frac{(\hat{a}^{\dagger }\hat{a}-\hat{b}%
^{\dagger }\hat{b}-\gamma l_{a})}{2}}  \notag \\
&&\times e^{i\phi _{+}\frac{(\hat{a}^{\dagger }\hat{a}+\hat{b}^{\dagger }%
\hat{b}-\gamma l_{a})}{2}}\eta _{a}^{\hat{n}_{a}/2}\hat{a}^{^{l_{a}}},
\end{eqnarray}%
where $\gamma =0$ and $\gamma =-1$ describe\ the photons loss before and
after the phase shifts, respectively.

According to Eqs. (20) and (26). The matrix elements $C_{ij}$ of Eq. (19)
can be worked out:
\begin{align}
C_{++}& =\left[ \eta _{a}+\gamma \eta _{a}-\gamma \right] ^{2}\langle \Delta
^{2}\hat{n}_{a}\rangle +\langle \Delta ^{2}\hat{n}_{b}\rangle  \notag \\
& +2\left[ \eta _{a}+\gamma \eta _{a}-\gamma \right] Cov[\hat{n}_{a},\hat{n}%
_{b}]  \notag \\
& +\left( \gamma +1\right) ^{2}(1-\eta _{a})\eta _{a}\langle \hat{n}%
_{a}\rangle , \\
C_{--}& =[\eta _{a}+\gamma \eta _{a}-\gamma ]^{2}\langle \Delta ^{2}\hat{n}%
_{a}\rangle +\langle \Delta ^{2}\hat{n}_{b}\rangle  \notag \\
& -2\left[ \eta _{a}+\gamma \eta _{a}-\gamma \right] Cov[\hat{n}_{a},\hat{n}%
_{b}]  \notag \\
& +\left( \gamma +1\right) ^{2}(1-\eta _{a})\eta _{a}\left\langle \hat{n}%
_{a}\right\rangle , \\
C_{+-}& =C_{-+}=[\eta _{a}+\gamma \eta _{a}-\gamma ]^{2}\langle \Delta ^{2}%
\hat{n}_{a}\rangle  \notag \\
& +\left( \gamma +1\right) ^{2}(1-\eta _{a})\eta _{a}\left\langle \hat{n}%
_{a}\right\rangle -\langle \Delta ^{2}\hat{n}_{b}\rangle .
\end{align}%
Substituting the matrix elements $C_{ij}$ into Eq.~(\ref{CQFIM}) and Eq.~(%
\ref{CQFIMP}), one can obtain the $\mathcal{C}_{QFIM_{\pm }}$ for the case
of loss in one arm, where $\mathcal{C}_{QFIM_{-}}$ and $\mathcal{C}%
_{QFIM_{+}}$ are SU(2) and SU(1,1) interferometers, respectively. They are
given as follows:
\begin{eqnarray}
\mathcal{C}_{QFIM_{-}} &=&4\frac{\Upsilon _{0}}{\Upsilon _{1}+2J\Upsilon _{2}%
},  \label{OptC_} \\
\text{ \ }\mathcal{C}_{QFIM+} &=&4\frac{\Upsilon _{0}}{\Upsilon
_{1}-2J\Upsilon _{2}},  \label{OptC+}
\end{eqnarray}%
where
\begin{eqnarray}
\Upsilon _{0} &=&\left\langle \hat{n}_{a}\right\rangle \left\langle \hat{n}%
_{b}\right\rangle (Q_{b}+1)[\left( \gamma +1\right) ^{2}(1-\eta _{a})\eta
_{a}  \notag \\
&&+(\eta _{a}+\gamma \eta _{a}-\gamma )^{2}(1-J^{2})(Q_{a}+1)],  \notag \\
\Upsilon _{1} &=&\left\langle \hat{n}_{a}\right\rangle (Q_{a}+1)(\eta
_{a}+\gamma \eta _{a}-\gamma )^{2}+\left\langle \hat{n}_{b}\right\rangle
(Q_{b}+1)  \notag \\
&&+\left( \gamma +1\right) ^{2}(1-\eta _{a})\eta _{a}\left\langle \hat{n}%
_{a}\right\rangle ,  \notag \\
\Upsilon _{2} &=&\left\langle \hat{n}_{a}\right\rangle (Q_{a}+1)(\eta
_{a}+\gamma \eta _{a}-\gamma )\sqrt{\frac{\left\langle \hat{n}%
_{b}\right\rangle (Q_{b}+1)}{\left\langle \hat{n}_{a}\right\rangle (Q_{a}+1)}%
}.\text{ \ }
\end{eqnarray}%
Next, we minimize $\mathcal{C}_{QFIM_{\pm }}$ by varying parameter $\gamma $%
, respectively.

\subsection{SU(2) interferometers}

In the case of SU(2) interferometers. To minimize the $\mathcal{C}%
_{QFIM_{-}} $, it requires to find the optimal $\gamma $, and we get%
\begin{equation}
\frac{d\mathcal{C}_{QFIM_{-}}}{d\gamma }=0.
\end{equation}%
After calculation, we obtain the optimal $\gamma _{opt}^{-}$, which is given
by
\begin{equation}
\gamma _{opt}^{-}=\frac{1}{(1-\eta _{a})+\frac{\eta _{a}}{(Q_{a}+1)(1-J^{2})}%
(1+J\sqrt{\frac{\left\langle \hat{n}_{a}\right\rangle (Q_{a}+1)}{%
\left\langle \hat{n}_{b}\right\rangle (Q_{b}+1)}})}-1.
\end{equation}%
Substituting the optimal $\gamma _{opt}^{-}$ into $\mathcal{C}_{QFIM_{-}}$of
Eq. (\ref{OptC_}), the optimal bound $\mathcal{C}_{QFIM_{-}}^{opt}$ is
worked out:
\begin{eqnarray}
&&\mathcal{C}_{QFIM_{-}}^{opt}=\frac{1}{\Upsilon _{3}}4(1-J^{2})[\langle
\hat{n}_{a}\rangle \frac{\eta _{a}}{1-\eta _{a}}(1-J^{2})\left\langle \hat{n}%
_{b}\right\rangle (Q_{b}+1)  \notag \\
&&+\langle \hat{n}_{a}\rangle ^{2}(\frac{\eta _{a}}{1-\eta _{a}})^{2}(\sqrt{%
\frac{\left\langle \hat{n}_{b}\right\rangle (Q_{b}+1)}{\left\langle \hat{n}%
_{a}\right\rangle (Q_{a}+1)}}+J)^{2}].  \label{CQ-opt}
\end{eqnarray}%
where
\begin{eqnarray}
\Upsilon _{3} &=&(\frac{\eta _{a}\langle \hat{n}_{a}\rangle }{1-\eta _{a}}%
)^{2}\{[\frac{1+5J^{2}}{\left\langle \hat{n}_{a}\right\rangle (Q_{a}+1)}+%
\frac{\left\langle \hat{n}_{b}\right\rangle (Q_{b}+1)}{\left\langle \hat{n}%
_{a}\right\rangle ^{2}(Q_{a}+1)^{2}}  \notag \\
&&+\frac{J^{2}}{\left\langle \hat{n}_{b}\right\rangle (Q_{b}+1)}+\frac{%
2J(1+J^{2})}{\sqrt{\left\langle \hat{n}_{a}\right\rangle \left\langle \hat{n}%
_{b}\right\rangle (Q_{a}+1)(Q_{b}+1)}}  \notag \\
&&+\frac{4J}{\left\langle \hat{n}_{a}\right\rangle (Q_{a}+1)}\sqrt{\frac{%
\left\langle \hat{n}_{b}\right\rangle (Q_{b}+1)}{\left\langle \hat{n}%
_{a}\right\rangle (Q_{a}+1)}}]  \notag \\
&&+\langle \hat{n}_{a}\rangle \frac{\eta _{a}}{1-\eta _{a}}[(1-J^{2})[1+2%
\frac{\left\langle \hat{n}_{b}\right\rangle (Q_{b}+1)}{\left\langle \hat{n}%
_{a}\right\rangle (Q_{a}+1)}  \notag \\
&&+J^{2}+4J\sqrt{\frac{\left\langle \hat{n}_{b}\right\rangle (Q_{b}+1)}{%
\left\langle \hat{n}_{a}\right\rangle (Q_{a}+1)}}]\}  \notag \\
&&+(1-J^{2})^{2}\left\langle \hat{n}_{b}\right\rangle (Q_{b}+1).
\end{eqnarray}

\begin{figure}[t]
\centerline{\includegraphics[width=0.5\textwidth,angle=0]{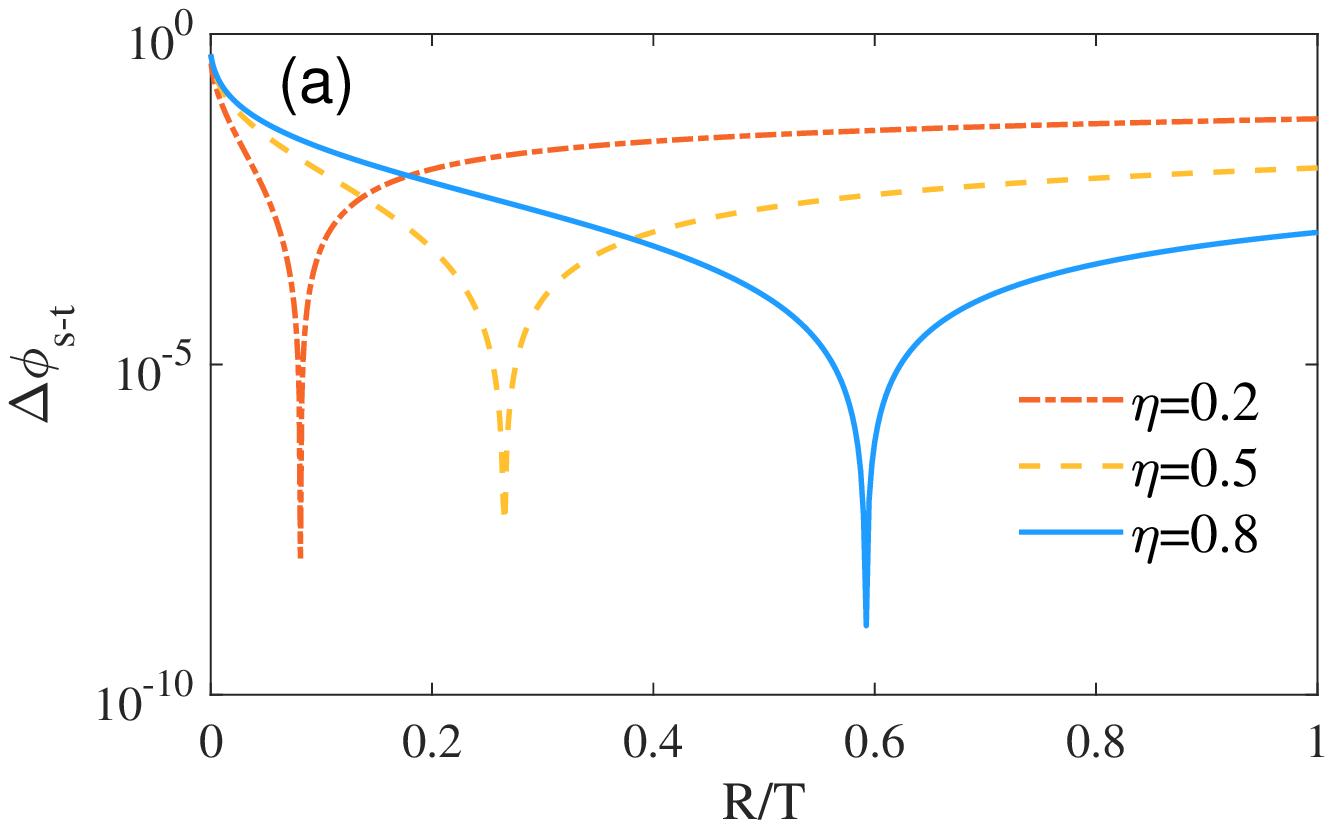}} %
\centerline{\includegraphics[width=0.5\textwidth,angle=0]{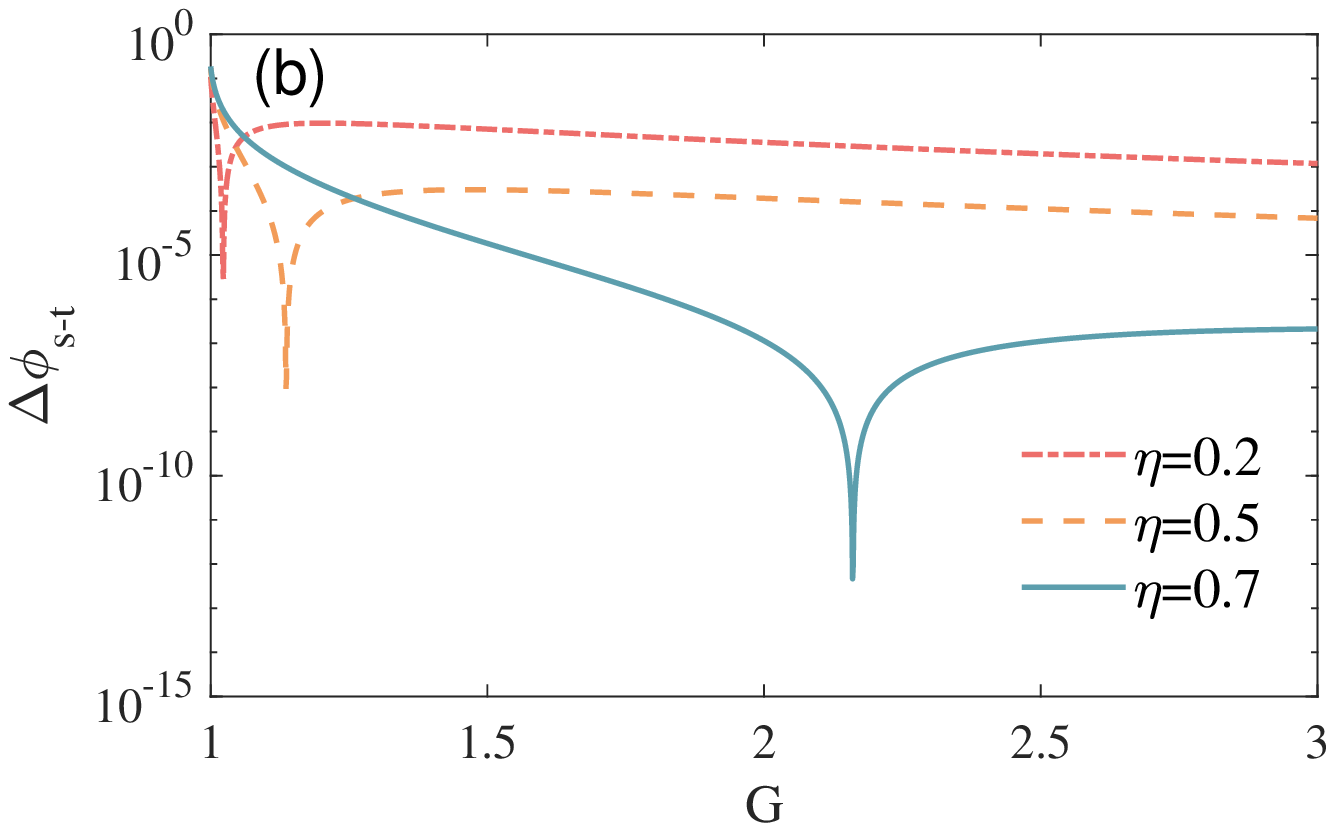}}
\caption{Phase sensitivity differences $\Delta \protect\phi _{s-t}$ between
the single-parameter estimation and the two-parameter estimation of (a)
SU(2) interferometer versus $R/T$ and (b) SU(1,1) interferometer versus $G$
under the condition of different photon loss coefficient $\protect\eta$,
where $|\protect\alpha|=2$, and $r=0.5$.}
\label{fig4}
\end{figure}

The bound $\mathcal{C}_{QFIM_{-}}^{opt}$ depends on the variance of the
initial state $\left\langle \Delta ^{2}\hat{n}_{a}\right\rangle $ and $%
\left\langle \Delta ^{2}\hat{n}_{b}\right\rangle $, the number of photons in
the initial state $\langle \hat{n}_{a}\rangle $, the intermode correlation $%
J $, and the losses $\eta _{a}$.

We check the optimal bound $\mathcal{C}_{QFIM_{-}}^{opt}$ in two limits. If
there is small dissipation, that is
\begin{equation}
\langle \Delta ^{2}\hat{n}_{b}\rangle ,\langle \Delta ^{2}\hat{n}_{a}\rangle
\ll \langle \hat{n}_{a}\rangle \frac{\eta _{a}}{1-\eta _{a}}.
\end{equation}%
Eq. (\ref{CQ-opt}) arrives at%
\begin{eqnarray}
\mathcal{C}_{QFIM_{-}}^{opt} &=&\frac{1}{\Upsilon _{3L}}[4(1-J^{2})\left%
\langle \hat{n}_{a}\right\rangle (Q_{a}+1)  \notag \\
&&\times (\sqrt{\frac{\left\langle \hat{n}_{b}\right\rangle (Q_{b}+1)}{%
\left\langle \hat{n}_{a}\right\rangle (Q_{a}+1)}}+J)^{2}],
\end{eqnarray}%
where
\begin{eqnarray}
\Upsilon _{3L} &=&1+J^{2}\frac{\left\langle \hat{n}_{a}\right\rangle
(Q_{a}+1)}{\left\langle \hat{n}_{b}\right\rangle (Q_{b}+1)}+\frac{%
\left\langle \hat{n}_{b}\right\rangle (Q_{b}+1)}{\left\langle \hat{n}%
_{a}\right\rangle (Q_{a}+1)}  \notag \\
&&+2J(J^{2}+1)\sqrt{\frac{\left\langle \hat{n}_{a}\right\rangle (Q_{a}+1)}{%
\left\langle \hat{n}_{b}\right\rangle (Q_{b}+1)}}+5J^{2}  \notag \\
&&+4J\sqrt{\frac{\left\langle \hat{n}_{b}\right\rangle (Q_{b}+1)}{%
\left\langle \hat{n}_{a}\right\rangle (Q_{a}+1)}}.
\end{eqnarray}

When $\left\langle \hat{n}_{a}\right\rangle =\left\langle \hat{n}%
_{b}\right\rangle =\bar{n}/2$, and $\langle \Delta ^{2}\hat{n}_{a}\rangle
=\langle \Delta ^{2}\hat{n}_{b}\rangle $, the overestimated Fisher
information tends to $0$, then the result reduces to lossless case $%
F_{QFI_{-}}=\bar{n}(Q+1)(1-J)$ \cite{Sahota15}.

In the opposite, highly dissipative limit
\begin{equation}
\langle \Delta ^{2}\hat{n}_{b}\rangle ,\langle \Delta ^{2}\hat{n}_{a}\rangle
\gg \langle \hat{n}_{a}\rangle \frac{\eta _{a}}{1-\eta _{a}},
\end{equation}%
Eq.~(\ref{CQ-opt}) is simplified as to
\begin{equation}
\mathcal{C}_{QFIM_{-}}^{opt}=\frac{\eta _{a}\langle \hat{n}_{a}\rangle }{%
1-\eta _{a}}(1-2J\frac{\langle \Delta \hat{n}_{b}\rangle }{\langle \Delta
\hat{n}_{a}\rangle })-\Delta \mathcal{C}_{QFIM_{-}},  \label{Eq41}
\end{equation}%
where the first term in $\mathcal{C}_{QFIM_{-}}$of Eq. (\ref{Eq41}) is the
bound from the single-parameter estimation, which coincides with the bound
by Escher \textit{et al.} \cite{Escher11} and Demkowicz-Dobrzanski \textit{%
et al. }\cite{Demkowicz}. The second term $\Delta \mathcal{C}_{QFIM_{-}}$ is
the overestimated Fisher information from QFI-only calculation in the
presence of losses in the SU(2) interferometers, which is given by
\begin{equation}
\Delta \mathcal{C}_{QFIM_{-}}=\frac{over_{U}^{-}}{over_{D}^{-}},
\end{equation}%
where
\begin{eqnarray}
over_{U}^{-} &=&\frac{\eta _{a}^{2}\langle \hat{n}_{a}\rangle ^{2}}{(1-\eta
_{a})^{2}}(1-2J\sqrt{\frac{\left\langle \hat{n}_{b}\right\rangle (Q_{b}+1)}{%
\left\langle \hat{n}_{a}\right\rangle (Q_{a}+1)}})  \notag \\
&&\times \lbrack 1-J^{2}+2(J+\sqrt{\frac{\left\langle \hat{n}%
_{b}\right\rangle (Q_{b}+1)}{\left\langle \hat{n}_{a}\right\rangle (Q_{a}+1)}%
})^{2}]  \notag \\
&&-\frac{\eta _{a}\langle \hat{n}_{a}\rangle }{1-\eta _{a}}%
(1-J^{2})\left\langle \hat{n}_{b}\right\rangle (Q_{b}+1)  \notag \\
&&\times (2J\sqrt{\frac{\left\langle \hat{n}_{b}\right\rangle (Q_{b}+1)}{%
\left\langle \hat{n}_{a}\right\rangle (Q_{a}+1)}}+3),  \notag \\
over_{D}^{-} &=&\frac{\eta _{a}\langle \hat{n}_{a}\rangle }{1-\eta _{a}}%
[1-J^{2}+2(J+\sqrt{\frac{\left\langle \hat{n}_{b}\right\rangle (Q_{b}+1)}{%
\left\langle \hat{n}_{a}\right\rangle (Q_{a}+1)}})^{2}]  \notag \\
&&+(1-J^{2})\left\langle \hat{n}_{b}\right\rangle (Q_{b}+1).
\end{eqnarray}%
When $\left\langle \hat{n}_{a}\right\rangle =\left\langle \hat{n}%
_{b}\right\rangle =\bar{n}/2$ and $\langle \Delta ^{2}\hat{n}_{a}\rangle
=\langle \Delta ^{2}\hat{n}_{b}\rangle $, the overestimated Fisher
information $\Delta \mathcal{C}_{QFIM_{-}}\neq 0$ in the presence of losses,
it is different from lossless case.

\subsection{SU(1,1) interferometers}

In the case of SU(1, 1) interferometers. To minimize the $\mathcal{C}%
_{QFIM_{+}}$, it requires to find the optimal $\gamma $, and we obtain%
\begin{equation}
\frac{d\mathcal{C}_{QFIM_{+}}}{d\gamma }=0.  \label{Gama}
\end{equation}%
Based on Eq. (\ref{Gama}), the optimal $\gamma _{opt}^{+}$ is obtained
\begin{equation}
\gamma _{opt}^{+}=\frac{1}{(1-\eta _{a})+\frac{\eta _{a}}{(Q_{a}+1)(1-J^{2})}%
(1-J\sqrt{\frac{\left\langle \hat{n}_{a}\right\rangle (Q_{a}+1)}{%
\left\langle \hat{n}_{b}\right\rangle (Q_{b}+1)}})}-1.
\end{equation}%
Substituting the optimal $\gamma _{opt}^{+}$ into $\mathcal{C}_{QFIM_{+}}$
of Eq. (\ref{OptC+}), the optimal bound $\mathcal{C}_{QFIM_{+}}^{opt}$ is
worked out:
\begin{eqnarray}
&&\mathcal{C}_{QFIM_{+}}^{opt}=\frac{4}{\Upsilon _{4}}[\langle \hat{n}%
_{a}\rangle \frac{\eta _{a}}{1-\eta _{a}}(1-J^{2})^{2}\left\langle \hat{n}%
_{b}\right\rangle (Q_{b}+1)  \notag \\
&&+\langle \hat{n}_{a}\rangle ^{2}(\frac{\eta _{a}}{1-\eta _{a}})^{2}(\sqrt{%
\frac{\left\langle \hat{n}_{b}\right\rangle (Q_{b}+1)}{\left\langle \hat{n}%
_{a}\right\rangle (Q_{a}+1)}}-J)^{2}(1-J^{2})],\text{ \ }
\end{eqnarray}%
where
\begin{eqnarray}
\Upsilon _{4} &=&(\frac{\eta _{a}\langle \hat{n}_{a}\rangle }{1-\eta _{a}}%
)^{2}\{[\frac{1+5J^{2}}{\left\langle \hat{n}_{a}\right\rangle (Q_{a}+1)}+%
\frac{\left\langle \hat{n}_{b}\right\rangle (Q_{b}+1)}{\left\langle \hat{n}%
_{a}\right\rangle ^{2}(Q_{a}+1)^{2}}  \notag \\
&&+\frac{J^{2}}{\left\langle \hat{n}_{b}\right\rangle (Q_{b}+1)}-\frac{%
2J(1+J^{2})}{\sqrt{\left\langle \hat{n}_{a}\right\rangle \left\langle \hat{n}%
_{b}\right\rangle (Q_{a}+1)(Q_{b}+1)}}  \notag \\
&&-\frac{4J}{\left\langle \hat{n}_{a}\right\rangle (Q_{a}+1)}\sqrt{\frac{%
\left\langle \hat{n}_{b}\right\rangle (Q_{b}+1)}{\left\langle \hat{n}%
_{a}\right\rangle (Q_{a}+1)}}]  \notag \\
&&+\langle \hat{n}_{a}\rangle \frac{\eta _{a}}{1-\eta _{a}}(1-J^{2})[1+2%
\frac{\left\langle \hat{n}_{b}\right\rangle (Q_{b}+1)}{\left\langle \hat{n}%
_{a}\right\rangle (Q_{a}+1)}  \notag \\
&&+J^{2}-4J\sqrt{\frac{\left\langle \hat{n}_{b}\right\rangle (Q_{b}+1)}{%
\left\langle \hat{n}_{a}\right\rangle (Q_{a}+1)}}]\}  \notag \\
&&+(1-J^{2})^{2}\left\langle \hat{n}_{b}\right\rangle (Q_{b}+1).
\end{eqnarray}

Similar to the SU(2) interferometers, we can also derive two limits. For
small dissipation, that is%
\begin{equation}
\langle \Delta ^{2}\hat{n}_{b}\rangle ,\langle \Delta ^{2}\hat{n}_{a}\rangle
\ll \langle \hat{n}_{a}\rangle \frac{\eta _{a}}{1-\eta _{a}}.
\end{equation}%
We get the reduced result
\begin{equation}
\mathcal{C}_{QFIM_{+}}^{opt}=4\frac{(\sqrt{\frac{\left\langle \hat{n}%
_{b}\right\rangle (Q_{b}+1)}{\left\langle \hat{n}_{a}\right\rangle (Q_{a}+1)}%
}-J)^{2}(1-J^{2})\left\langle \hat{n}_{a}\right\rangle (Q_{a}+1)}{\Upsilon
_{4L}},
\end{equation}%
where
\begin{eqnarray}
\Upsilon _{4L} &=&1+5J^{2}+J^{2}\frac{\left\langle \hat{n}_{a}\right\rangle
(Q_{a}+1)}{\left\langle \hat{n}_{b}\right\rangle (Q_{b}+1)}+\frac{%
\left\langle \hat{n}_{b}\right\rangle (Q_{b}+1)}{\left\langle \hat{n}%
_{a}\right\rangle (Q_{a}+1)}  \notag \\
&&-2J(1+J^{2})\sqrt{\frac{\left\langle \hat{n}_{a}\right\rangle (Q_{a}+1)}{%
\left\langle \hat{n}_{b}\right\rangle (Q_{b}+1)}}  \notag \\
&&-4J\sqrt{\frac{\left\langle \hat{n}_{b}\right\rangle (Q_{b}+1)}{%
\left\langle \hat{n}_{a}\right\rangle (Q_{a}+1)}}.
\end{eqnarray}

When $\left\langle \hat{n}_{a}\right\rangle =\left\langle \hat{n}%
_{b}\right\rangle =\bar{n}/2$, and $\langle \Delta ^{2}\hat{n}_{a}\rangle
=\langle \Delta ^{2}\hat{n}_{b}\rangle $, the overestimated Fisher
information tends to $0$, then the result reduces to lossless case $%
F_{QFI_{+}}=\bar{n}(Q+1)(1+J)$ \cite{Gong17}. Unlike the case of SU(2)
interferometers, here is $1+J$ not $1-J$.

The highly dissipative limit is%
\begin{equation}
\langle \Delta ^{2}\hat{n}_{b}\rangle ,\langle \Delta ^{2}\hat{n}_{a}\rangle
\gg \langle \hat{n}_{a}\rangle \frac{\eta _{a}}{1-\eta _{a}}.
\end{equation}%
we obtain%
\begin{equation}
\mathcal{C}_{QFIM_{+}}^{opt}=\frac{\eta _{a}\langle \hat{n}_{a}\rangle }{%
1-\eta _{a}}(1+2J\frac{\langle \Delta \hat{n}_{b}\rangle }{\langle \Delta
\hat{n}_{a}\rangle })-\Delta \mathcal{C}_{QFIM_{+}},
\end{equation}%
where
\begin{equation}
\Delta \mathcal{C}_{QFIM_{+}}=\frac{over_{U}^{+}}{over_{D}^{+}},
\end{equation}%
with%
\begin{eqnarray}
over_{U}^{+} &=&\frac{\eta _{a}^{2}\langle \hat{n}_{a}\rangle ^{2}}{(1-\eta
_{a})^{2}}(1+2J\sqrt{\frac{\left\langle \hat{n}_{b}\right\rangle (Q_{b}+1)}{%
\left\langle \hat{n}_{a}\right\rangle (Q_{a}+1)}})  \notag \\
&&\times \lbrack 1-J^{2}+2(J-\sqrt{\frac{\left\langle \hat{n}%
_{b}\right\rangle (Q_{b}+1)}{\left\langle \hat{n}_{a}\right\rangle (Q_{a}+1)}%
})^{2}]  \notag \\
&&+\frac{\eta _{a}\langle \hat{n}_{a}\rangle }{1-\eta _{a}}%
(1-J^{2})\left\langle \hat{n}_{b}\right\rangle (Q_{b}+1)  \notag \\
&&\times (2J\sqrt{\frac{\left\langle \hat{n}_{b}\right\rangle (Q_{b}+1)}{%
\left\langle \hat{n}_{a}\right\rangle (Q_{a}+1)}}-3),  \notag \\
over_{D}^{+} &=&\frac{\eta _{a}\langle \hat{n}_{a}\rangle }{1-\eta _{a}}%
[1-J^{2}+2(J-\sqrt{\frac{\left\langle \hat{n}_{b}\right\rangle (Q_{b}+1)}{%
\left\langle \hat{n}_{a}\right\rangle (Q_{a}+1)}})^{2}]  \notag \\
&&+(1-J^{2})\left\langle \hat{n}_{b}\right\rangle (Q_{b}+1).
\end{eqnarray}%
Similar to SU(2) interferometer case, when $\left\langle \hat{n}%
_{a}\right\rangle =\left\langle \hat{n}_{b}\right\rangle =\bar{n}/2$ and $%
\langle \Delta ^{2}\hat{n}_{a}\rangle =\langle \Delta ^{2}\hat{n}_{b}\rangle
$, the overestimated Fisher information of $\Delta \mathcal{F}_{+}=0$ in
lossless case becomes $\Delta \mathcal{C}_{QFIM_{-}}\neq 0$ in the presence
of losses. Compare these two expressions $\Delta \mathcal{C}_{QFIM_{+}}$ and
$\Delta \mathcal{C}_{QFIM-}$, there is only difference in signs between
them. Since LBS and NBS are linear and nonlinear processes respectively, the
average photon number and photon fluctuation of the same input state after
LBS and NBS transformation will have significant differences, thus the phase
sensitivity will also have significant differences.

\section{Losses in two arms}
Interferometers with photon losses in both arms as shown in
Fig. \ref{fig1} can be treated in a similar way. A possible set of Kraus
operators describing the process is
\begin{align}
\hat{\Pi}_{l_{a},l_{b}}& =\sqrt{\frac{(1-\eta _{a})^{l_{a}}}{l_{a}!}}\sqrt{%
\frac{(1-\eta _{b})^{l_{b}}}{l_{b}!}}  \notag \\
\times e^{i\phi _{-}\frac{\hat{n}_{a}-\hat{n}_{b}-\gamma _{a}l_{a}+\gamma
_{b}l_{b}}{2}}& e^{i\phi _{+}\frac{\hat{n}_{a}+\hat{n}_{b}-\gamma
_{a}l_{a}-\gamma _{b}l_{b}}{2}}\eta _{a}^{\frac{\hat{n}_{a}}{2}}\eta _{b}^{%
\frac{\hat{n}_{b}}{2}}\hat{a}^{l_{a}}\hat{b}^{l_{b}},
\end{align}%
where $\eta _{a}$ ($\eta _{b}$) quantifies the photon losses of arm $a$ ($b$%
). $\gamma _{a}=-1$ and $\gamma _{b}=-1$ ($\gamma _{a}=0$ and $\gamma _{b}=0$%
) describe\ the photons loss before (after) the phase shifts of arm $a$ and
arm $b$.

Substituting the $\hat{\Pi}_{l_{a},l_{b}}(\phi _{+},\phi _{-})$ into $C_{ij}$%
\ of Eq.~(\ref{CQ}), the matrix elements $C_{ij}$ are given as follows:
\begin{eqnarray}
&&C_{++}=[1-\Gamma _{a}(1-\eta _{a})]^{2}\left\langle \Delta \hat{n}%
_{a}^{2}\right\rangle +\Gamma _{a}^{2}(1-\eta _{a})\eta _{a}\left\langle
\hat{n}_{a}\right\rangle  \notag \\
&&+[1-\Gamma _{b}(1-\eta _{b})]^{2}\left\langle \Delta \hat{n}%
_{b}^{2}\right\rangle +\Gamma _{b}^{2}(1-\eta _{b})\eta _{b}\left\langle
\hat{n}_{b}\right\rangle  \notag \\
&&+2[1-\Gamma _{a}(1-\eta _{a})][1-\Gamma _{b}(1-\eta _{b})]Cov[\hat{n}_{a},%
\hat{n}_{b}],
\end{eqnarray}%
\begin{eqnarray}
&&C_{--}=[1-\Gamma _{a}(1-\eta _{a})]^{2}\left\langle \Delta \hat{n}%
_{a}^{2}\right\rangle +\Gamma _{a}^{2}(1-\eta _{a})\eta _{a}\left\langle
\hat{n}_{a}\right\rangle  \notag \\
&&+[1-\Gamma _{b}(1-\eta _{b})]^{2}\left\langle \Delta \hat{n}%
_{b}^{2}\right\rangle +\Gamma _{b}^{2}(1-\eta _{b})\eta _{b}\left\langle
\hat{n}_{b}\right\rangle  \notag \\
&&-2[1-\Gamma _{a}(1-\eta _{a})][1-\Gamma _{b}(1-\eta _{b})]Cov[\hat{n}_{a},%
\hat{n}_{b}],
\end{eqnarray}%
\begin{eqnarray}
&&C_{+-}=C_{-+}=[1-\Gamma _{a}(1-\eta _{a})]^{2}\left\langle \Delta \hat{n}%
_{a}^{2}\right\rangle +\Gamma _{a}^{2}(1-\eta _{a})\left\langle \hat{n}%
_{a}\right\rangle  \notag \\
&&\times \eta _{a}-[1-\Gamma _{b}(1-\eta _{b})]^{2}\left\langle \Delta \hat{n%
}_{b}^{2}\right\rangle -\Gamma _{b}^{2}(1-\eta _{b})\eta _{b}\left\langle
\hat{n}_{b}\right\rangle ,
\end{eqnarray}%
where $\Gamma _{a}=\gamma _{a}+1$, $\Gamma _{b}=\gamma _{b}+1$.

Substituting the matrix elements $C_{ij}$ into $\mathcal{C}_{QFIM_{-}}$\ of
Eq. (\ref{CQFIM}) and into $\mathcal{C}_{QFIM_{+}}$\ of Eq. (\ref{CQFIMP}),
we can obtain the $\mathcal{C}_{QFIM_{-}}$ and $\mathcal{C}_{QFIM_{+}}$. For
convenience, considering $\gamma _{a}=\gamma _{b}=\gamma \equiv \Omega -1$, $%
\eta _{a}=\eta _{b}=\eta $, $\mathcal{C}_{QFIM_{-}}$ and $\mathcal{C}%
_{QFIM_{+}}$ are given by
\begin{equation}
\mathcal{C}_{QFIM_{-}}=4\frac{\Upsilon _{5}}{[1-\Omega (1-\eta
)]_{-}^{2}\chi _{-}+\Omega ^{2}(1-\eta )\eta \epsilon },
\end{equation}%
and
\begin{equation}
\mathcal{C}_{QFIM_{+}}=4\frac{\Upsilon _{5}}{[1-\Omega (1-\eta )]^{2}\chi
_{+}+\Omega ^{2}(1-\eta )\eta \epsilon },
\end{equation}%
where
\begin{eqnarray}
\Upsilon _{5} &=&[1-\Omega (1-\eta )]^{4}\zeta +\Omega ^{4}(1-\eta )^{2}\eta
^{2}\tau  \notag \\
&&+[1-\Omega (1-\eta )]^{2}\Omega ^{2}(1-\eta )\eta \lambda ,
\end{eqnarray}%
and%
\begin{eqnarray}
\chi _{\pm } &=&\left\langle \hat{n}_{a}\right\rangle (Q_{a}+1)+\left\langle
\hat{n}_{b}\right\rangle (Q_{b}+1)  \notag \\
&&\mp 2J\sqrt{\left\langle \hat{n}_{a}\right\rangle \left\langle \hat{n}%
_{b}\right\rangle (Q_{a}+1)(Q_{b}+1)},  \notag \\
\zeta &=&(1-J^{2})\left\langle \hat{n}_{a}\right\rangle \left\langle \hat{n}%
_{b}\right\rangle (Q_{a}+1)(Q_{b}+1),  \notag \\
\epsilon &=&\left\langle \hat{n}_{a}\right\rangle +\left\langle \hat{n}%
_{b}\right\rangle ,\tau =\left\langle \hat{n}_{a}\right\rangle \left\langle
\hat{n}_{b}\right\rangle ,  \notag \\
\lambda &=&\left\langle \hat{n}_{a}\right\rangle \left\langle \hat{n}%
_{b}\right\rangle \left[ (Q_{a}+1)+(Q_{b}+1)\right] .
\end{eqnarray}

Similar to single-arm loss case, we also minimize $\mathcal{C}_{QFIM_{-}}$
and $\mathcal{C}_{QFIM_{+}}$, corresponding to SU(2) and SU(1,1)
interferometers, respectively.

\subsection{SU(2) interferometers}

In the case of SU(2) interferometers. To minimize the $\mathcal{C}%
_{QFIM_{-}} $, it requires to find the optimal $\gamma $, and we get%
\begin{equation}
\frac{d\mathcal{C}_{QFIM_{-}}}{d\gamma }=0.
\end{equation}%
The equation above is more complex, and we cannot obtain an analytical
solution. Therefore, we check the $d\mathcal{C}_{QFIM_{-}}/d\gamma $
directly\ in two limits again. If there is small dissipation, we recover the
QFIM $\mathcal{F}_{QFIM_{-}}$ for the lossless case.

In the opposite, if there is highly dissipative limit, due to high intensity
we have $\left\langle \Delta \hat{n}_{b}^{2}\right\rangle _{0}\approx
\left\langle \Delta \hat{n}_{a}^{2}\right\rangle _{0}$, and consider a
special case $J=-1$ then we obtain the optimal $\gamma _{opt}^{H-}$, which
is given by
\begin{equation}
\gamma _{opt}^{H-}=\frac{\eta \left[ I-\tau \right] }{\left[ \eta \tau
+(1-\eta )\lambda \right] },
\end{equation}%
where the superscript $H$ indicates high loss. Substituting the optimal $%
\gamma _{opt}^{H-}$ into Eq. (\ref{CQFIMP}), the optimal bound $\mathcal{C}%
_{QFIM_{+}}^{H}$ is worked out:
\begin{eqnarray}
&&\mathcal{C}_{QFIM_{-}}^{H}=4\frac{1}{(\Lambda _{-}^{H})^{2}\chi
_{-}+(\Omega _{-}^{H})^{2}(1-\eta )\eta \epsilon }[(\Lambda
_{-}^{H})^{4}\zeta  \notag \\
&&+(\Omega _{-}^{H})^{4}(1-\eta )^{2}\eta ^{2}\tau +(\Lambda
_{-}^{H})^{2}(\Omega _{-}^{H})^{2}(1-\eta )\eta \lambda ],
\end{eqnarray}%
where%
\begin{equation}
\Omega _{-}^{H}=\frac{\lambda }{\left[ \eta \tau +(1-\eta )\lambda \right] },%
\text{ }\Lambda _{-}^{H}=\frac{\eta \tau }{\left[ \eta \tau +(1-\eta
)\lambda \right] }.
\end{equation}

\subsection{SU(1,1) interferometers}

Substituting the matrix elements $C_{ij}$ into Eq. (\ref{CQFIM}) and Eq. (%
\ref{CQFIMP}), ones can obtain the $\mathcal{C}_{QFIM_{-}}$ and $\mathcal{C}%
_{QFIM_{+}}$.

To minimize the $\mathcal{C}_{QFIM_{+}}$, it requires to find the optimal $%
\gamma $, and we obtain
\begin{equation}
\frac{d\mathcal{C}_{QFIM_{+}}}{d\gamma }=0.
\end{equation}

Similar to the case of SU(2) interferometers, due to the above equation is
more complex, we check the $d\mathcal{C}_{QFIM_{+}}/d\gamma $ directly\ in
two limits again. If there is small dissipation, we recover the QFIM $%
\mathcal{F}_{QFIM+}$ for the lossless case.

In the opposite, if there is highly dissipative limit, due to high intensity
we have $\left\langle \Delta \hat{n}_{b}^{2}\right\rangle _{0}\approx
\left\langle \Delta \hat{n}_{a}^{2}\right\rangle _{0}$, and we also consider
a special case $J=1$, we obtain the optimal $\gamma _{opt}^{H+}$, which is
given by
\begin{equation}
\gamma _{opt}^{H+}=\frac{\eta \left[ I-\tau \right] }{\left[ \eta \tau
+(1-\eta )\lambda \right] },
\end{equation}%
Substituting the optimal $\gamma _{opt}^{H+}$ into Eq. (\ref{CQFIMP}), the
optimal bound $\mathcal{C}_{QFIM_{+}}^{H}$ is cast into:
\begin{eqnarray}
&&\mathcal{C}_{QFIM_{+}}^{H}=4\frac{1}{(\Lambda _{+}^{H})^{2}\chi
_{+}+(\Omega _{+}^{H})^{2}(1-\eta )\eta \epsilon }[(\Lambda
_{+}^{H})^{4}\zeta  \notag \\
&&+(\Lambda _{+}^{H}\Omega _{+}^{H})^{2}(1-\eta )\eta \lambda +(\Omega
_{+}^{H})^{4}(1-\eta )^{2}\eta ^{2}\tau ],
\end{eqnarray}%
where
\begin{equation}
\Omega _{+}^{H}=\frac{\lambda }{\left[ \eta \tau +(1-\eta )\lambda \right] },%
\text{ }\Lambda _{+}^{H}=\frac{\eta \tau }{\left[ \eta \tau +(1-\eta
)\lambda \right] }.
\end{equation}

\section{QCRB and Numerical results}

The phase sensitivity of the interferometer can be obtained for a given
measurement scheme, such as homodyne measurement \cite{Li14}, parity
measurement \cite{Li16} or intensity measurement \cite{Plick10}, with usage
of error propagation formula. However, it is difficult to optimize over the
detection methods to obtain the optimal estimation schemes. Fortunately, the
QFI introduced by Braustein and Caves \cite{Braunstein94,Braunstein96} is
the intrinsic information in the quantum state and is not related to a
particular measurement scheme. Based on the QFI, the ultimate precision
bound of phase sensitivity is given by the QCRB
\begin{equation}
\Delta \phi _{QCRB}=\frac{1}{\sqrt{m\mathcal{F}}},
\end{equation}%
where $m$ is the number of independent repeats of the experience.

Now we present a typical example to apply our theory to describe the
difference between the QFI-only and QFIM. Here, We consider the case that
the two input modes of the interferometer. We input the $\left\vert \alpha
\right\rangle _{a}\otimes \left\vert \varsigma \right\rangle _{b}$, where
coherent state $\left\vert \alpha \right\rangle $ with $\alpha =\left\vert
\alpha \right\vert e^{i\theta _{\alpha }}$ with $\left\vert \alpha
\right\vert $ being a complex number and $\theta _{\alpha }$ being the
initial phase. And the squeezed vacuum state $\left\vert \varsigma
\right\rangle =\exp [1/2(\varsigma ^{\ast }\hat{a}^{2}-\varsigma (\hat{a}%
^{\dagger })^{2})$ with $\varsigma =re^{i\theta _{r}}$\ is squeezed
parameter where $r$ and $\theta _{r}$ are the squeezing amplitude and
squeezing angle respectively.

\subsection{SU(2) interferometers}

In the SU(2) interferometer, the average number of photons in the two arms
is $\langle \hat{n}_{a}\rangle ^{-}=T\left\vert \alpha \right\vert
^{2}+R\sinh ^{2}r$ and $\langle \hat{n}_{b}\rangle ^{-}=R\left\vert \alpha
\right\vert ^{2}+T\sinh ^{2}r$, when $2\theta _{\alpha }-\theta _{r}=0$, we
can get
\begin{eqnarray}
\langle \Delta ^{2}\hat{n}_{a}\rangle ^{-} &=&T^{2}\left\vert \alpha
\right\vert ^{2}+2R^{2}\sinh ^{2}r\cosh ^{2}r  \notag \\
&&+TR(\left\vert \alpha \right\vert ^{2}e^{2r}+\sinh ^{2}r),  \notag \\
\langle \Delta ^{2}\hat{n}_{b}\rangle ^{-} &=&R^{2}\left\vert \alpha
\right\vert ^{2}+2T^{2}\sinh ^{2}r\cosh ^{2}r  \notag \\
&&+TR(\left\vert \alpha \right\vert ^{2}e^{2r}+\sinh ^{2}r),
\end{eqnarray}%
and%
\begin{equation}
Cov[\hat{n}_{a},\hat{n}_{b}]^{-}=TR(\left\vert \alpha \right\vert
^{2}(1-e^{2r})+\sinh ^{2}r\cosh 2r).
\end{equation}%
where $T$ and $R$ are the reflectivity and transmissivity of the BS,
respectively. And using the above results one can get%
\begin{eqnarray*}
Q_{a}^{-} &=&\frac{R(T\left\vert \alpha \right\vert ^{2}(e^{2r}-1)+R\sinh
^{2}r\cosh 2r)}{T\left\vert \alpha \right\vert ^{2}+R\sinh ^{2}r}, \\
Q_{b}^{-} &=&\frac{T(R\left\vert \alpha \right\vert ^{2}(e^{2r}-1)+T\sinh
^{2}r\cosh 2r)}{R\left\vert \alpha \right\vert ^{2}+T\sinh ^{2}r}.
\end{eqnarray*}%
\begin{equation}
J^{-}=\frac{TR(\left\vert \alpha \right\vert ^{2}(1-e^{2r})+\sinh ^{2}r\cosh
2r)}{\sqrt{%
\begin{array}{c}
\left[
\begin{array}{c}
T\left\vert \alpha \right\vert ^{2}(T+Re^{2r}) \\
+R\sinh ^{2}r(T+2R\cosh ^{2}r)%
\end{array}%
\right] \\
\times \left[
\begin{array}{c}
R\left\vert \alpha \right\vert ^{2}(R+Te^{2r}) \\
+T\sinh ^{2}r(R+2T\cosh ^{2}r)%
\end{array}%
\right]%
\end{array}%
}}.
\end{equation}

\subsection{SU(1,1) interferometers}

Similar to the SU(2) interferometer case, we also consider the input $%
\left\vert \alpha \right\rangle _{a}\otimes \left\vert \varsigma
\right\rangle _{b}$ in the SU(1,1) interferometer. The average number of
photons in the two arms is $\langle \hat{n}_{a}\rangle ^{+}=G^{2}\left\vert
\alpha \right\vert ^{2}+g^{2}\cosh ^{2}r$ and $\langle \hat{n}_{b}\rangle
^{+}=G^{2}\sinh ^{2}r+g^{2}(\left\vert \alpha \right\vert ^{2}+1)$, when $%
2\theta _{g}-2\theta _{\alpha }-\theta _{r}=\pi $, $\theta _{g}$ is the
phase shift of the NBS for wave splitting and recombination. According to
the above conditions, we obtain%
\begin{eqnarray}
\langle \Delta ^{2}n_{a}\rangle ^{+} &=&G^{4}\left\vert \alpha \right\vert
^{2}+2g^{4}\sinh ^{2}r\cosh ^{2}r  \notag \\
&&+G^{2}g^{2}(\left\vert \alpha \right\vert ^{2}e^{2r}+\cosh ^{2}r),  \notag
\\
\langle \Delta {}^{2}n_{b}\rangle ^{+} &=&g^{4}\left\vert \alpha \right\vert
^{2}+2G^{4}\sinh ^{2}r\cosh ^{2}r  \notag \\
&&+G^{2}g^{2}(\left\vert \alpha \right\vert ^{2}e^{2r}+\cosh ^{2}r),
\end{eqnarray}%
and%
\begin{equation}
Cov[\hat{n}_{a},\hat{n}_{b}]^{+}=G^{2}g^{2}(\left\vert \alpha \right\vert
^{2}(1+e^{2r})+\cosh ^{2}r\cosh 2r).
\end{equation}

Here, $G$ is the gain factors of NBS, for wave splitting and recombination
with $G^{2}-g^{2}=1$. The results are consistent with the results of Ref.~%
\cite{You19}. Using the above results we work out%
\begin{eqnarray}
Q_{a}^{+} &=&\frac{g^{2}(G^{2}\left\vert \alpha \right\vert
^{2}(1+e^{2r})+g^{2}\cosh ^{2}r\cosh 2r)}{G^{2}\left\vert \alpha \right\vert
^{2}+g^{2}\cosh ^{2}r},  \notag \\
Q_{b}^{+} &=&\frac{\left[
\begin{array}{c}
G^{2}\sinh ^{2}r\left[ 2G^{2}\cosh ^{2}r+g^{2}-1\right] \\
+g^{2}(g^{2}(\left\vert \alpha \right\vert ^{2}+1)+\left\vert \alpha
\right\vert ^{2}(G^{2}e^{2r}-1))%
\end{array}%
\right] }{G^{2}\sinh ^{2}r+g^{2}(\left\vert \alpha \right\vert ^{2}+1)}.
\end{eqnarray}%
\begin{equation}
J^{+}=\frac{G^{2}g^{2}(\left\vert \alpha \right\vert ^{2}(1+e^{2r})+\cosh
^{2}r\cosh 2r)}{\sqrt{%
\begin{array}{c}
\left[
\begin{array}{c}
G^{2}\left\vert \alpha \right\vert ^{2}(G^{2}+g^{2}e^{2r}) \\
+g^{2}\cosh ^{2}r(G^{2}+2g^{2}\sinh ^{2}r)%
\end{array}%
\right] \\
\times \left[
\begin{array}{c}
g^{2}\left\vert \alpha \right\vert ^{2}(g^{2}+G^{2}e^{2r}) \\
+G^{2}\cosh ^{2}r(g^{2}+2G^{2}\sinh ^{2}r)%
\end{array}%
\right]%
\end{array}%
}}.
\end{equation}

Using the above results, we can obtain the numerical QCRBs in the case of
losses in one arm and in two arms with the coherent state and squeezed
vacuum state input for an example.

Firstly, we study the phase sensitivities of SU(1,1) interferometers or
SU(2) interferometers and compare the differences between the results given
by the QFI and QFIM phase estimation methods without loss. As shown in Fig.~%
\ref{fig2}, the phase sensitivities of the SU(1,1) interferometer or SU(2)
interferometer has difference between the single-parameter estimation and
the two-parameter estimation under certain conditions. The difference is due
to $\Delta F_{\pm }$ also known as overestimated Fisher Information. With
the increase of $N_{\alpha }$, the difference of phase sensitivity caused by
$\Delta F_{\pm }$ always exists, which corresponds to our conclusion in Sec.~%
\ref{sec:IQ}. Therefore, it is appropriate to use the two-parameter
estimation for the phase estimation of the interferometer.

Then, in the presence of loss of one arm, we study the phase sensitivities
of SU(1,1) interferometers or SU(2) interferometers in the method of QFIM,
as is depicted in Fig.~\ref{fig3}. The QCRBs of the single-parameter
estimation and the two-parameter estimation are very close, the sensitivity
of them is approximately equal at moderate loss. In order to investigate
this phenomenon, we study the difference of phase sensitivity $\Delta \phi
_{s-t}$ between single-parameter and two-parameter estimation by changing $%
\eta $. In the case of SU(2) interferometers, $\Delta \phi _{s-t}$ as a
function of $R/T$ under different photon loss coefficients $\eta $ is shown
in Fig.~\ref{fig4}(a). For a given $\eta $, there is a minimum value of $%
\Delta \phi _{s-t}$, which appears on the larger beam splitter ratio $R/T$
as $\eta $ increases. In the case of SU(1,1) ones, $\Delta \phi _{s-t}$ as a
function of $G$ under different photon loss coefficients $\eta $ is shown in
Fig.~\ref{fig4}(b). Similar to the SU(2) case, there is also a minimum value
of $\Delta \phi _{s-t}$, which appears on the larger $G$ as $\eta $
increases.

The overestimated Fisher Information in the ideal case gradually disappears
as the loss increases. With the further increase of the loss, the
overestimated Fisher Information reappears. The area of these minimum values
of $\Delta \phi _{s-t}$ corresponds to $\Delta C_{\pm }=0$, i.e., $%
C_{+-}=C_{-+}=0$. Due to consider the single-arm loss (arm $a$), Eq.
(29) can be written as $C_{+-}=C_{-+}=[\eta _{a}+\gamma \eta _{a}-\gamma ]^{2}\langle
\Delta ^{2}\hat{n}_{a}\rangle +\left( \gamma +1\right) ^{2}(1-\eta _{a})\eta
_{a}\left\langle \hat{n}_{a}\right\rangle -\langle \Delta ^{2}\hat{n}%
_{b}\rangle \equiv \overline{\langle \Delta ^{2}\hat{n}_{a}\rangle }-\langle
\Delta ^{2}\hat{n}_{b}\rangle $, where $\overline{\langle \Delta ^{2}\hat{n}%
_{a}\rangle }$ can be seen as effective fluctuation due to loss. Because the
fluctuations $\overline{\langle \Delta ^{2}\hat{n}_{a}\rangle }$ and $%
\langle \Delta ^{2}\hat{n}_{b}\rangle $ are different, when the loss
increases gradually, $\overline{\langle \Delta ^{2}\hat{n}_{a}\rangle }$
decreases gradually. When it decreases to the same as the other arm, the
overestimated Fisher Information disappears. The loss continues to increase,
and the fluctuation $\overline{\langle \Delta ^{2}\hat{n}_{a}\rangle }$ in
turn continues to increase, leading to the re-emergence of overestimation. A
similar phenomenon also occurs in the loss of both arms.

\section{conclusion}

In conclusion, we theoretically extended the model developed by Escher
\textit{et al} \cite{Escher11} to the QFIM case with noise. Photon loss is a
very usual noise in optical systems, then we gave the QFIM expressions in the
case of single-arm photon loss and two-arm photon losses in the SU(2) and
SU(1,1) interferometers with arbitrary pure states input. The ultimate
precision limits of SU(2) and SU(1,1) interferometers with photon losses was
investigated and discussed by using coherent state $\otimes $ squeezed
vacuum state as an example. The overestimated Fisher Information existing in
the ideal case will occur disappear and revival phenomenon with loss
coefficient or splitter ratio changing. For a given loss coefficient, adjusting the splitter ratio is a method to optimize the sensitivity of quantum measurements in a lossy environment. This strategy is beneficial to quantum precision measurement in lossy environments.

\section{Acknowledgments}

This work is supported by the National Natural Science Foundation of China
Grants No.~11974111, No.~11874152, No.~12104423, No.~11654005, No. 11974116,
No.~91536114, Shanghai Municipal Science and Technology Major Project under
Grant No. 2019SHZDZX01; Innovation Program of Shanghai Municipal Education
Commission No. 202101070008E00099; the Fundamental Research Funds for the
Central Universities; the Shanghai talent program.

\end{document}